\documentclass[preprint,aps,prd,epsfig] {revtex4}
\pdfoutput=1
\usepackage{epsfig}
\usepackage{color}
\usepackage{bm}
\usepackage{subfigure}
\usepackage{leftidx}

\newcommand{\ba}{\begin{array}}
\newcommand{\ea}{\end{array}}

\begin{document}
\draft
\title{{\bf $X(1835)$ and the New Resonances $X(2120)$ and $X(2370)$ Observed by the BES Collaboration}}

\author{Jia-Feng Liu, Gui-Jun Ding and Mu-Lin Yan}

\affiliation{\centerline{Department of Modern
Physics,}\centerline{University of Science and Technology of
China,Hefei, Anhui 230026, China}}

\begin{abstract}

We calculate the decay widths of both the second and the third
radial excitations of $\eta$ and $\eta'$ within the framework of
$^3P_0$ model. After comparing the theoretical decay widths and
decay patterns with the available experimental data of $\eta(1760)$,
$X(1835)$, $X(2120)$ and $X(2370)$, we find that the interpretation
of $\eta(1760)$ and $X(1835)$ as the second radial excitation of
$\eta$ and $\eta'$ crucially depends on the measured mass and width
of $\eta(1760)$, which is still controversial experimentally. We
suggest that there may be sizable $p\bar{p}$ content in $X(1835)$.
$X(2120)$ and $X(2370)$ can not be understood as the third radial
excitations of $\eta$ and $\eta'$, $X(2370)$ probably is a mixture
of $\eta'(4{^1}{S}{_0})$ and glueball.

\vskip0.5cm


\end{abstract}
\maketitle
\section{introduction}

$X(1835)$ was first observed by BESII in the $\eta'\pi\pi$ invariant
mass spectrum in the process
$J/\psi\rightarrow\gamma\pi^+\pi^-\eta'$ with a statistical
significance of 7.7$\sigma$. The fit with the Breit-Wigner function
yields mass $M=1833.7\pm 6.1(stat)\pm 2.7(syst) \rm{MeV}/c^2$, width
$\Gamma=67.7\pm20.3(stat)\pm7.7(syst) \rm{MeV}/c^2$ and the product
branching fraction $Br(J/\psi\rightarrow\gamma
X(1835))Br(X(1835)\rightarrow\pi^{+}\pi^{-}\eta^{\prime})
=(2.2\pm0.4(stat)\pm0.4(syst))\times10^{-4}$ \cite{Ablikim:2005um}.
Recently X(1835) has been confirmed by BESIII collaboration in the
same process with statistical significance larger than 25$\sigma$,
and its mass and width are fitted to be $M=1838.1\pm2.8$ MeV and
$\Gamma=179.5\pm9.1$ MeV. Moreover two new resonances are reported,
which are denoted as $X(2120)$ and $X(2370)$ respectively. Their
masses and widths are determined to be $M_{X(2120)}=2124.8\pm5.6$
MeV, $\Gamma_{X(2120)}=101\pm14$ MeV, $M_{X(2370)}=2371.0\pm6.4$ MeV
and $\Gamma_{X(2370)}=108\pm15$ MeV \cite{ichep_Yuan,ichep_Huang}.

The experimental observation of $X(1835)$ stimulated a number of
theoretical speculations about its underlying structure. Some
interpret $X(1835)$ as a $p\overline{p}$ bound state
\cite{Datta:2003iy,Ding:2005ew,Zhu:2005ns,Chang:2004us,Loiseau:2005cv},
a glueball candidate
\cite{Kochelev:2005vd,He:2005nm,Li:2005vd,Hao:2005hu} or the radial
excitation of $\eta'$ \cite{Huang:2005bc,Klempt:2007cp}, and some
others interpret it as final state interaction or a rescattering
effect \cite{Zou:2003zn,Chen:2008ee,Liu:2009vm}. Naively the
observation of $X(2120)$ and $X(2370)$ seems to indicate that all
the three resonances $X(1835)$, $X(2120)$ and $X(2370)$ are possibly
the radial excitations of $\eta$ or $\eta'$, they jump to the ground
state $\eta'$ through emitting two $\pi$ \footnote{This conjecture
is proposed in Ref. \cite{ichep_Yuan} as well.}. Moreover, we note
that before we consider the exotic structure hypothesis for some
newly observed resonance, it is very necessary to study whether the
assignment of conventional hadron is possible. Consequently, we
shall investigate in the following whether $X(1835)$, $X(2120)$ and
$X(2370)$ could be canonical $q\bar{q}$ pseudoscalar mesons.

It is well-known that there are nine pseudoscalar mesons $\pi$, $K$,
$\eta$ and $\eta'$, which form a good nonet in the limit of SU(3)
flavor symmetry. From Particle Data Group(PDG) \cite{pdg}, we see
that the first radial excitations of these pseudoscalars have been
well established, concretely they are $\pi(1300)$, $K(1460)$,
$\eta(1295)$ and $\eta(1475)$. As a result, if the three resonances
$X(1835)$, $X(2120)$ and $X(2370)$ are canonical $q\bar{q}$
pseudoscalar mesons, the natural assignment would be $\eta(1760)$
and $X(1835)$ as the second radial excitation of $\eta$ and $\eta'$,
$X(2120)$ and $X(2370)$ as the third radial excitation of $\eta$ and
$\eta'$ respectively. In this work, we shall study the decays of
these four resonances under the above assignment within the
framework of $^3P_0$ model. Our goal is to shed some light on the
nature of these structures by comparing the predictions for the
hadronic decay widths with the available experimental data.

This paper is organized as follows. Firstly we review the
${^3}{P}{_0}$ model briefly in sections II. The flavor mixing
between the $\eta$ and $\eta'$ radial excitation and the allowed
decay modes are presented in section III. The OZI (Okubo, Zweig and
Iizuka) allowed strong decays of $\eta(1760)$, $X(1835)$, $X(2120)$
and $X(2370)$ are studied in section IV. Finally we present in
section V our conclusions and some discussions.


\section{Review of the ${^3}{P}{_0}$ model}
\begin{figure}[t]
\vspace{-0.3cm}
\begin{center}
\makebox{\epsfxsize=5.0in\epsffile{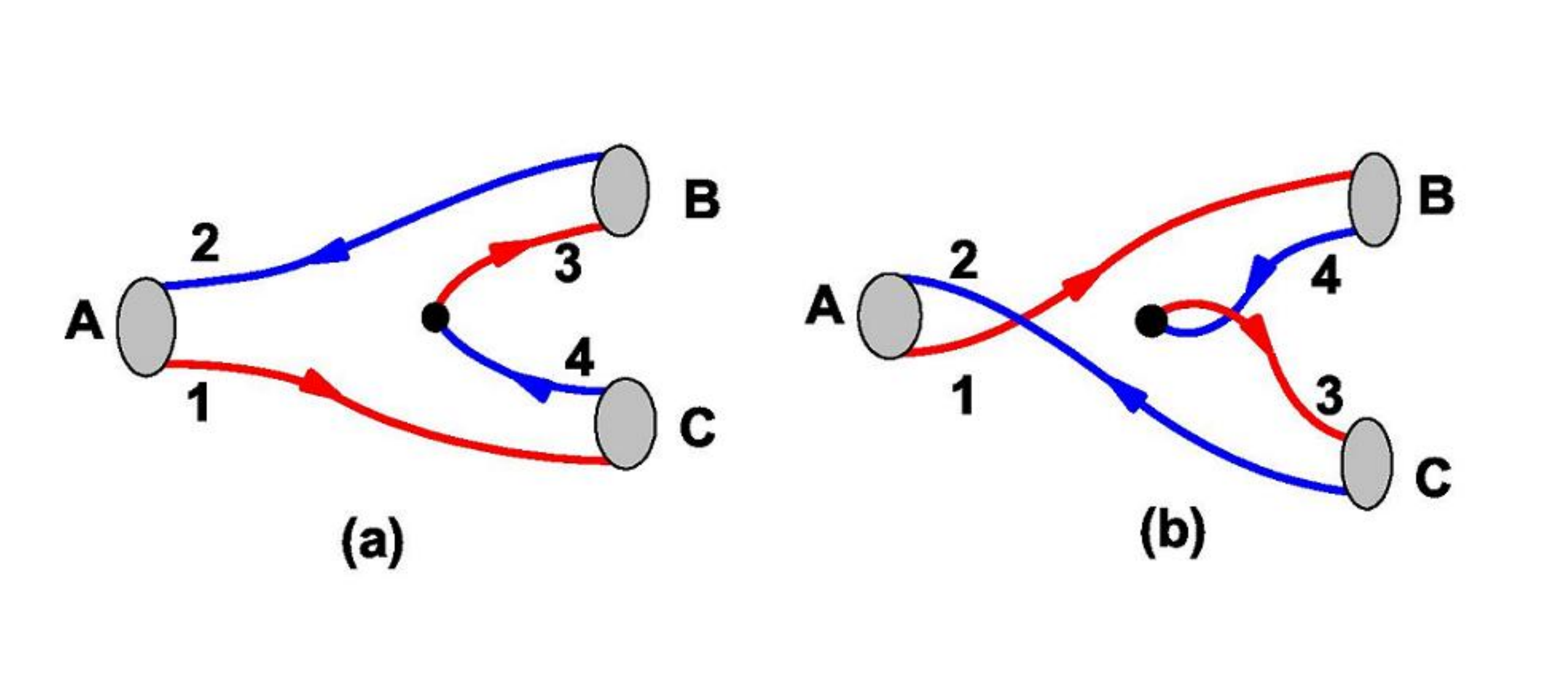}}
\end{center}
\vspace{-0.5cm} \caption{\label{fig0}Two possible diagrams
contributing to the meson decay $A \rightarrow B
C$ in the $^3P_0$ model.} 
\end{figure}

The ${^3}{P}{_0}$ model for the decay of a $q\overline{q}$ meson $A$
to mesons $B+C$ was proposed by Micu\cite{Micu:1969a} and developed
by Le Yaouanc et al \cite{orsay,Geiger:1994kr,Ackleh:1996yt}. The
${^3}{P}{_0}$ model assumes that strong decay takes place via the
creation of a pair of quark and antiquark with $J^{PC}=0^{++}$ from
the vacuum. The created quark pair together with the quark and
antiquark in the initial meson recombine to final state mesons in
two ways as shown in Fig.\ref{fig0}, and the decay amplitude is
proportional to wavefunctions (including spatial, spin, flavor and
color wavefunctions) overlap between the initial state, created
quark pair and the final state. The ${^3}{P}{_0}$ model has been
widely applied to meson and baryon strong decays, with considerable
success
\cite{orsay,Geiger:1994kr,Ackleh:1996yt,Barnes:1996ff,Barnes:2002mu,Godfrey:1986wj,Barnes:2005pb,Capstick:2000qj}.
In this work, we shall use the diagrammatic technique developed in
Ref. \cite{Ackleh:1996yt} to derive the amplitudes and the $^3P_0$
matrix elements. In this formalism, the $^3P_0$ model describes the
strong decay process using a $q\bar{q}$ pair production Hamiltonian,
which is the nonrelativistic limit of,
\begin{equation}
\label{1}H_I=g\int d^3\mathbf{x}\;\overline{\psi}(x)\;\psi(x)
\end{equation}
where $\psi$ is a Dirac quark field, $g$ is the coupling constant.
The pair production component of the $^3P_0$ Hamiltonian $H_I$ can
be written in terms of creation operators as
\begin{equation}
\label{2} H_{I}=\sum_{s\overline{s}}\int d^{3}\mathbf{k} \frac{g
m_{q}}{E_{q}}[\overline{u}_{\mathbf{k}s}v_{\mathbf{-k}\overline{s}}]b^{\dagger}_{\mathbf{k}s}d^{\dagger}_{\mathbf{-k}\bar{s}}
\end{equation}
where $b^{\dagger}_{\mathbf{k}s}$ creates a quark with momentum
$\mathbf{k}$ and spin $s$, $d^{\dagger}_{\mathbf{-k}\bar{s}}$ create
a antiquark with momentum $\mathbf{-k}$ and spin $\bar{s}$, $m_q$
being the mass of the created quark and antiquark. We note that each
effective $^3P_0$ quark pair production vertex is associated with
the factor $\frac{g
m_{q}}{E_{q}}[\overline{u}_{\mathbf{k}s}v_{\mathbf{-k}\overline{s}}]$.
We assume non-relativistic $q\bar{q}$ wavefunction for the initial
and final mesons,
\begin{equation}
\label{3}\mid A\rangle=\int d^{3}\mathbf{p_1}\int
d^{3}\mathbf{p_2}\;\Psi_{n_AL_AM_{L_A}}(\frac{m_2\mathbf{p_1}-m_1\mathbf{p_2}}{m_1+m_2})\delta(\mathbf{P_A}-\mathbf{p_1}-\mathbf{p_2})\mid
q_1(\mathbf{p_1})\bar{q}_2(\mathbf{p_2})>
\end{equation}
with explicit spin and flavor wave functions which are of the usual
non-relativistic quark model forms. $n_A$ denotes the radial quantum
number of meson $A$ composed of quark $q_1$ and anti-quark
$\bar{q}_2$ with momentum $\mathbf{p_1}$ and $\mathbf{p_2}$ and mass
$m_{1}$ and $m_{2}$ respectively, and $\mathbf{P_A}$ is the momentum
of meson $A$. The wavefunctions of the final state mesons $B$ and
$C$ can be written out directly in the same way. The spatial
wavefunction $\Psi$ is generally taken to be the simple harmonic
oscillator (SHO) wavefunction.  The SHO wavefunction enables
analytical calculation of the decay amplitude, and it turned out to
be a good approximation. Even if we use more realistic wavefunction,
the predictions would not be improved systematically due to the
inherent uncertainties of the $^3P_0$ model. In momentum-space, the
SHO wavefunction reads
\begin{equation}
\label{4}\Psi_{nLM_L}(\mathbf{p})=\frac{(-1)^n(-i)^L}{\beta^{3/2}}\sqrt{\frac{2n!}{\Gamma(n+L+3/2)}}\Big(\frac{p}{\beta}\Big)^L\exp\Big(-\frac{p^2}{2\beta^2}\Big)L^{L+1/2}_n\Big(\frac{p^2}{\beta^2}\Big)Y_{LM_L}(\Omega_p)
\end{equation}
where $\beta$ is the harmonic oscillator parameter,
$Y_{LM_L}(\Omega_p)$ is the spherical harmonic function, and
$L^{L+1/2}_n\Big(\frac{p^2}{\beta^2}\Big)$ is the Laguerre
polynomial.

One can now straightforwardly evaluate the Hamiltonian $H_I$ matrix
element for the decay $A\rightarrow B+C$ in terms of overlap
integrals,
\begin{eqnarray}
\nonumber&&\langle
BC|H_I|A\rangle_{a}=I_{signature}(a)I_{flavor}(a)I_{spin+space}(a)\delta(\mathbf{P_A}-\mathbf{P_B}-\mathbf{P_C})\\
\label{5}&&\langle
BC|H_I|A\rangle_{b}=I_{signature}(b)I_{flavor}(b)I_{spin+space}(b)\delta(\mathbf{P_A}-\mathbf{P_B}-\mathbf{P_C})
\end{eqnarray}
where the signature phase $I_{signature}$ is equal to -1 for both
diagrams $(a)$ and $(b)$ due to quark operator anticommutation.
Starting from the flavor wavefunctions, we can directly obtain the
flavor overlap factors $I_{flavor}(a)$ and $I_{flavor}(b)$ which
result from contracting the explicit flavor states corresponding to
diagrams Fig.\ref{fig0}a and Fig.\ref{fig0}b, they are listed in
Table \ref{tab:flavor} for the decay modes concerned here. In the
rest frame of meson $A$, the overlap integral $I_{spin+space}(a)$
and $I_{spin+space}(b)$ explicitly are given by
\begin{eqnarray}
\nonumber I_{spin+space}(a)&=&\int
d^3\mathbf{k}\;\Psi_{n_AL_AM_{L_A}}(\mathbf{k}-\mathbf{P_B})\Psi^*_{n_BL_BM_{L_B}}(\mathbf{k}-\frac{m_3}{m_2+m_3}\mathbf{P_B})\Psi^*_{n_CL_CM_{L_C}}(\mathbf{k}-\frac{m_3}{m_1+m_3}\mathbf{P_B})\\
\nonumber&&\times
g\frac{m_3}{E_3}[\bar{u}_{\mathbf{k}s_{q_3}}v_{\mathbf{-k}s_{\bar{q}_4}}]\\
\nonumber I_{spin+space}(b)&=&\int
d^3\mathbf{k}\;\Psi_{n_AL_AM_{L_A}}(\mathbf{k}+\mathbf{P_B})\Psi^*_{n_BL_BM_{L_B}}(\mathbf{k}+\frac{m_3}{m_1+m_3}\mathbf{P_B})\Psi^*_{n_CL_CM_{L_C}}(\mathbf{k}+\frac{m_3}{m_2+m_3}\mathbf{P_B})\\
\label{6}&&\times
g\frac{m_3}{E_3}[\bar{u}_{\mathbf{k}s_{q_3}}v_{\mathbf{-k}s_{\bar{q}_4}}]
\end{eqnarray}
where the relevant spin factor has been omitted,
$E_3=\sqrt{\mathbf{k}^2+m^2_3}$ is the energy of the created quark.
We note that the spin factor and the labels $s_{q_3}$ and
$s_{\bar{q}_4}$ depend on the reaction considered, generally the
spin indexes $s_{q_3}$ and $s_{\bar{q}_4}$ associated with diagram
Fig.\ref{fig0}a and Fig.\ref{fig0}b are different. As a result, the
amplitude for the meson decay $A\rightarrow B+C$ is
\begin{equation}
\label{7}{\cal M}(A\rightarrow
B+C)=I_{signature}(a)I_{flavor}(a)I_{spin+space}(a)+I_{signature}(b)I_{flavor}(b)I_{spin+space}(b)\equiv
h_{fi}
\end{equation}
Taking into account the phase space, we get the differential decay
rate
\begin{equation}
\label{8}\frac{d\Gamma_{A\rightarrow BC}}{d\Omega}=2\pi\frac{P
E_{B}E_{C}}{M_{A}}|h_{fi}|^{2}
\end{equation}
where $E_B$ and $E_C$ are the energy of the meson $B$ and $C$
respectively, $P$ is the momentum of the final state mesons in the
rest frame of meson $A$
\begin{equation}
\label{9}P=\sqrt{[M^2_A-(M_B+M_C)^2][M^2_A-(M_B-M_C)^2]}\Big/(2M_A)
\end{equation}
where $M_A$, $M_B$ and $M_C$ are the masses of the meson $A$, $B$
and $C$ respectively. 
To compare with experiments, we transform the amplitude $h_{fi}$
into the partial wave amplitude ${\cal M}_{LS}$ by the recoupling
calculation \cite{Jacob:1959at}, then the decay width is
\begin{equation}
\label{add2}\Gamma(A\rightarrow B+C)=2\pi\frac{P
E_{B}E_{C}}{M_{A}}\sum_{LS}|{\cal M}_{LS}|^2
\end{equation}


The pair production parameter $g$ and the harmonic oscillator
parameter $\beta$ are fitted to the strong decay data, and they are
found to be roughly flavor independent for decays involving
production of $u\bar{u}$, $d\bar{d}$ and $s\bar{s}$ pairs. The
typical values obtained from computation of light meson decays are
$g=0.334$ GeV and $\beta=0.4$ GeV
\cite{Ackleh:1996yt,Barnes:1996ff,Barnes:2002mu}, assuming simple
harmonic oscillator wavefunctions with a global scale, and they are
frequently adopted by the literatures. However, different quark
models find different values of $\beta$ (mostly in the range of
$0.35\sim0.45$ GeV ), so that there is the question of the
sensitivity of our results to $\beta$, we will address this issue
below. The masses of constituent quarks are chosen to be
$m_u=m_d=0.33$ GeV and $m_s=0.55$ GeV as usual. The masses used are
the experimental values of well-established candidates, which are
taken from the PDG \cite{pdg}. Moreover, we have ignored the mass
difference between the members of the same isospin multiplet. For
the isoscalar we assume ideal mixing
$|\varphi_{nonstrange}\rangle=1/\sqrt{2}|u\bar{u}+d\bar{d}\rangle$,
$|\varphi_{strange}\rangle=|s\bar{s}\rangle$, where except for the
ground state pseudoscalar, we choose
$|\eta\rangle=\cos\phi_p|u\bar{u}+d\bar{d}\rangle/\sqrt{2}-\sin\phi_p
|s\bar{s}\rangle$ and
$|\eta'\rangle=\sin\phi_p|u\bar{u}+d\bar{d}\rangle/\sqrt{2}+\cos\phi_p
|s\bar{s}\rangle$ with the mixing angle $\phi_p=39.2^{\circ}$
\cite{Bramon:1997va}. The kaons and their excitations are not charge
conjugation eigenstates so that mixing can occur among states with
the same $J^P$ that are forbidden for neutral states. For example
the $J^P=1^+$ axial vector kaon mesons $K_1(1273)$ and $K_1(1402)$
are coherent superpositions of quark model $^3P_1$ and $^1P_1$
states \cite{Barnes:2002mu},
\begin{eqnarray}
\nonumber&&|K_{1}(1273)\rangle=\sqrt{\frac{2}{3}}\;|{^1}{P}{_1}\rangle+\sqrt{\frac{1}{3}}\;|{^3}{P}{_1}\rangle\\
\label{add1}&&|K_{1}(1402)\rangle=-\sqrt{\frac{1}{3}}\;|{^1}{P}{_1}\rangle+\sqrt{\frac{2}{3}}\;|{^3}{P}{_1}\rangle
\end{eqnarray}

\section{Mixing between the $\eta$ and $\eta'$ excitations and the allowed decay modes}
The radial excitation of $\eta$ and $\eta'$ are both isoscalar
states with the same $J^{PC}$ so that there will be mixing between
them. Consequently the physical states are the mixture of $SU(3)$
flavor octet and singlet
\begin{eqnarray}
\nonumber&&|\eta(n^1S_0)\rangle=\cos\theta|\eta_8(n^1S_0)\rangle-\sin\theta|\eta_0(n^1S_0)\rangle\\
\label{10}&&|\eta'(n^1S_0)\rangle=\sin\theta|\eta_8(n^1S_0)\rangle+\cos\theta|\eta_0(n^1S_0)\rangle
\end{eqnarray}
where $n$ represents the radial quantum number,
$|\eta_8(n^1S_0)\rangle$ and $|\eta_0(n^1S_0)\rangle$ are the octet
and singlet states respectively,
\begin{eqnarray}
\nonumber&&|\eta_8(n^1S_0)\rangle\equiv\frac{1}{\sqrt{6}}|u\bar{u}+d\bar{d}-2s\bar{s}\rangle\\
\label{11}&&|\eta_0(n^1S_0)\rangle\equiv\frac{1}{\sqrt{3}}|u\bar{u}+d\bar{d}+s\bar{s}\rangle
\end{eqnarray}
In order to explicitly exhibit the $u\bar{u}+d\bar{d}$ and
$s\bar{s}$ components, we shall choose the so-called
nonstrange-strange basis in this work
\begin{eqnarray}
\nonumber&&|\eta(n^1S_0)\rangle=\cos\phi|\eta_{NS}(n^1S_0)\rangle-\sin\phi|\eta_S(n^1S_0)\rangle\\
\label{12}&&|\eta'(n^1S_0)\rangle=\sin\phi|\eta_{NS}(n^1S_0)\rangle+\cos\phi|\eta_S(n^1S_0)\rangle
\end{eqnarray}
where $|\eta_{NS}(n^1S_0)\rangle=|u\bar{u}+d\bar{d}\rangle/\sqrt{2}$
and $|\eta_S(n^1S_0)\rangle=|s\bar{s}\rangle$, and mixing angle
$\phi$ is related to $\theta$ via
$\phi=\theta+\arctan\sqrt{2}\simeq\theta+54.7^\circ$. We note that
the mixing angle $\phi$ (or $\theta$) is less constrained
phenomenologically, its concrete value has to be determined
experimentally. It is well-known that $\eta-\eta'$ mixing has been
measured by various means, however, there is still large
uncertainty. As a result, we shall take the mixing angle $\phi$ as a
undetermined parameter in the following, the dependence of the
amplitudes and widths on $\phi$ would be considered.

We present the selection rules for the two-body decays of $\eta$ and
$\eta'$ excitations in Table \ref{tab:decaymodes}. For specific
final states listed in Table \ref{tab:decaymodes}, all the four
states $\eta(1760)$, $X(1835)$, $X(2120)$ and $X(2370)$ could decay
into them, if the process is not forbidden kinetically. We note that
decays into two pseudoscalar or two scalar mesons are forbidden by
parity and charge conjugation conservation, Moreover, the $G-$parity
forbids the decay processes $X\rightarrow\rho\pi$,
$X\rightarrow\omega\eta$, $X\rightarrow\rho a_1(1260)$,
$X\rightarrow\rho a_2(1320)$, $X\rightarrow\omega(\phi)f_1(1285)$,
$X\rightarrow\omega(\phi)f_1(1420)$,
$X\rightarrow\omega(\phi)f_2(1270)$ and
$X\rightarrow\omega(\phi)f'_2(1525)$, where $X$ denotes
$\eta(1760)$, $X(1835)$, $X(2120)$ or $X(2370)$.

\begin{table}[hptb]
\begin{center}
\renewcommand{\arraystretch}{0.5}
\begin{tabular}{cc} \hline\hline
Decay modes&~~~~~Final states\\\hline
$X\rightarrow {1\;^1S_0}+{1\;^3S_1}$&~~~~~$KK^*$\\
$X\rightarrow {2\;^1S_0}+{1\;^3S_1}$&~~~~~$K(1460)K^*$\\
$X\rightarrow {1\;^1S_0}+{2\,^3S_1}$&~~~~~$KK^*(1410)$\\
$X\rightarrow {1\;^1S_0}+{1\;^3P_0}$&~~~~~$\pi a_0(1450),\;KK^{*}_0(1430),\;\eta f_0(1370),\;\eta f_0(1710),\;\eta' f_0(1370)$\\
$X\rightarrow {1\;^1S_0}+{1\;^3P_2}$&~~~~~$\pi a_2(1320),\;KK^{*}_2(1430),\;\eta f_2(1270),\;\eta f'_2(1525),\;\eta' f_2(1270)$\\
$X\rightarrow {1\;^1S_0}+{1\;^3D_1}$&~~~~~$KK^*(1680)$\\
$X\rightarrow {1\;^1S_0}+{1\;^3D_3}$&~~~~~$KK^*_3(1780)$\\
$X\rightarrow {1\;^3S_1}+{1\;^3S_1}$&~~~~~$\rho\rho,\;K^*K^*,\;\omega\omega,\;\phi\phi$\\
$X\rightarrow {1\;^1S_0}+{2\;^3S_1}$&~~~~~$\rho\rho(1450),\;K^*K^*(1410),\;\omega\omega(1420)$\\
$X\rightarrow {1\;^3S_1}+{1\;^3P_1}$&~~~~~$\rho b_1(1235),\;K^*K_1(1273),\;\omega h_1(1170),\;\omega h_1(1380),\;\phi h_1(1170),\;\phi h_1(1380)$\\
$X\rightarrow {1\;^3S_1}+{1\;^3P_1}$&~~~~~$K^*K_1(1402)$\\
$X\rightarrow
{1\;^3S_1}+{1\;^3P_2}$&~~~~~$K^*K_2^*(1430)$\\\hline\hline
\end{tabular}
\caption{\label{tab:decaymodes} Allowed decay modes of $\eta$ and
$\eta'$ radial excitations.}
\end{center}
\end{table}

\section{Strong decays of $\eta(1760)$, $X(1835)$, $X(2120)$ and $X(2370)$ }
Following the procedures presented in the previous sections, the
total decay rate is given by the Hamilton matrix element squared,
multiplied by the phase space, and summed over all final spin and
charge states. Since we neglect mass splitting within the isospin
multiplet, to sum over all channels, one should multiply the partial
width into the specific charge channel by the flavor multiplicity
factor ${\cal F}$ in Table \ref{tab:flavor}. This ${\cal F}$ factor
also incorporates the statistical factor $1/2$ if the final state
mesons $B$ and $C$ are identical.


\subsection{Decays of $\eta(1760)$ and $X(1835)$ }
\begin{figure}[hptb]
\begin{center}
\begin{tabular}{cc}
\includegraphics[height=50mm]{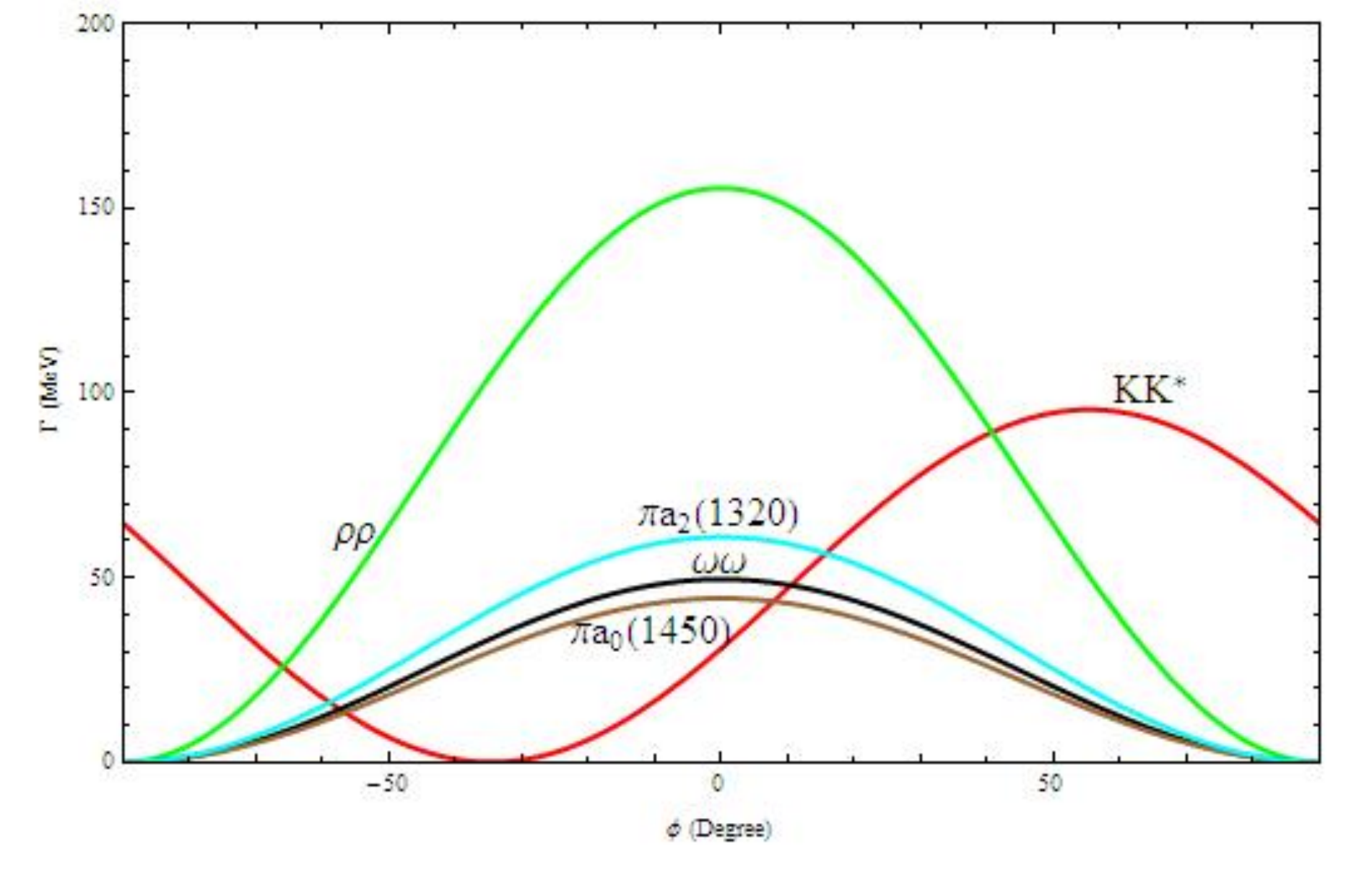}&\hspace{1cm}\includegraphics[height=50mm]{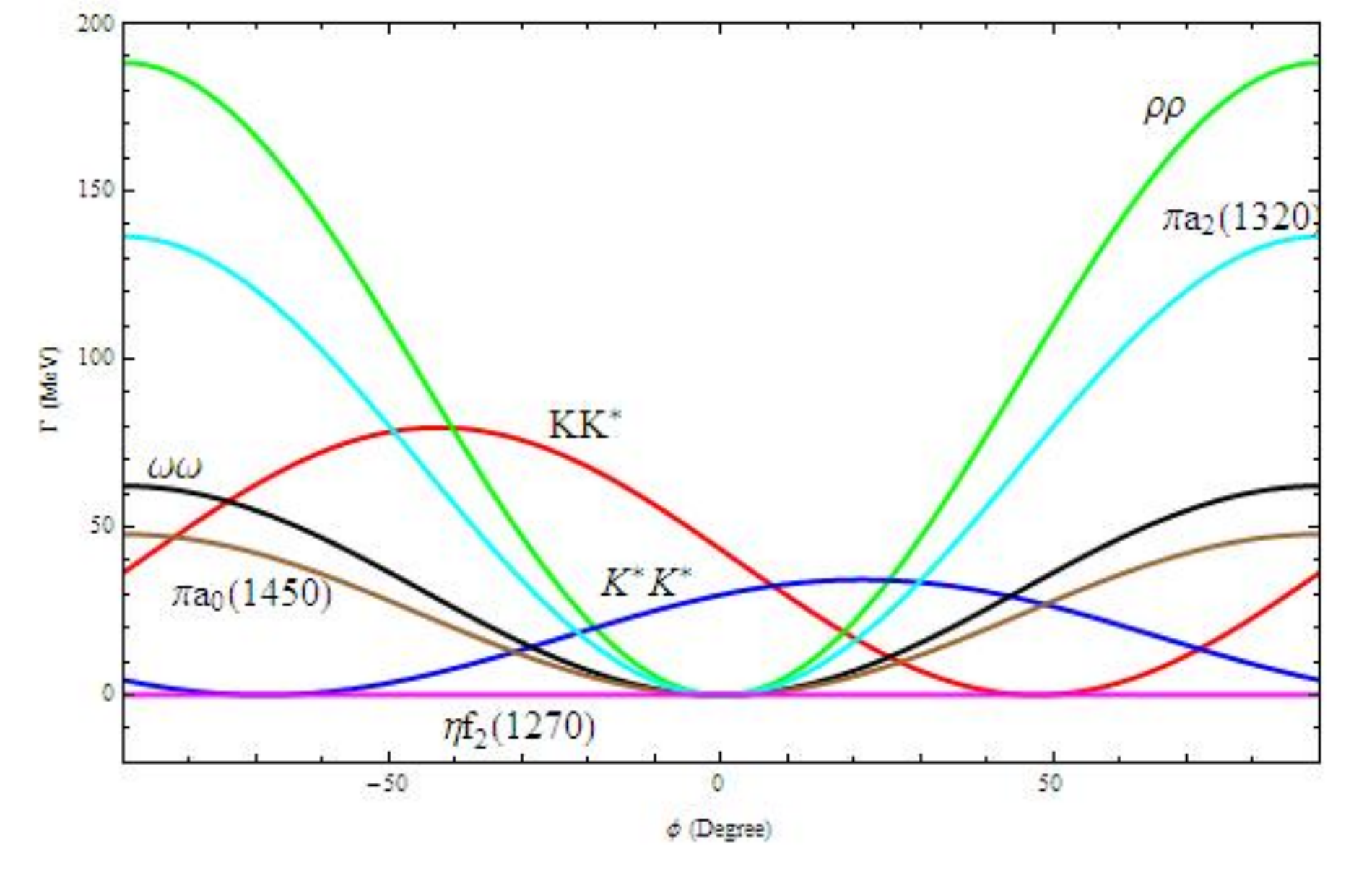}\\
(a)&(b)
\end{tabular}
\caption{\label{fig:x1835_eta1760_partial_widths} Partial decay
widths of $\eta(1760)$ and $X(1835)$ vs. the flavor mixing angle
$\phi$, the left figure for $\eta(1760)$, and the right figure for
$X(1835)$.}
\end{center}
\end{figure}

The experimental evidence for $\eta(1760)$ is controversial, its
existence evidence was first reported by the Mark III Collaboration
in the $J/\psi$ radiative decays to $\omega\omega$
\cite{Baltrusaitis:1985zi} and $\rho\rho$
\cite{Baltrusaitis:1985nd}, then it was further studied by the DM2
and BES collaborations. The various experimental results associated
with $\eta(1760)$ are summarized in Table \ref{tab:eta1760ex}, it is
obviously that there are big differences between different
measurements of $\eta(1760)$ width. In this work, both the mass and
width are taken to be the world average listed in PDG.
\begin{table}[hptb]
\begin{center}
\renewcommand{\arraystretch}{0.5}
\begin{tabular}{cccc} \hline\hline
Experiment&~~ Mass (MeV) &~~~ Width (MeV) &~~~ Production
\\\hline
DM2\cite{Bisello:1988as}& $1760\pm 11$ &~~~ $60\pm16$
&~~~$J/\psi\rightarrow
\gamma\eta(1760),~\eta(1760)\rightarrow\rho\rho$\\
BES\cite{Ablikim:2006ca}&$1744\pm10\pm15$ &~~~$244^{+24}_{-21}\pm25$
&~~~ $J/\psi\rightarrow
\gamma\eta(1760),~\eta(1760)\rightarrow\omega\omega$\\
PDG \cite{pdg}& $1756\pm9$ &~~~$96\pm70$ & \qquad \\\hline\hline
\end{tabular}
\caption{\label{tab:eta1760ex} Summary of $\eta(1760)$
measurements.}
\end{center}
\end{table}
For $\eta(1760)$ and $X(1835)$ as the second radial excitation of
$\eta$ and $\eta'$, the allowed decay channels, the corresponding
decay amplitudes and partial widths are shown in Table
\ref{tab:eta1760} and Table \ref{tab:x1835} respectively. Clearly
the decay amplitudes and widths depend strongly on the mixing angle
$\phi$, and measurements of any or several of the larger decay modes
will provide constrained tests of the hypothesis and measurement of
the mixing angle. We believe that the better way to determine the
mixing angle is comparing the ratio between $KK^*$ and $\rho\rho$
partial widths with experimental data, if both $\eta(1760)$ and
$X(1835)$ are indeed conventional quark model states assumed above.
This is because that the pair production parameter $g$ cancels out
in this ratio, consequently there is less systematic uncertainty
than in the decay rates. The partial widths of $\eta(1760)$ and
$X(1835)$ as functions of the flavor mixing angle $\phi$ for fixed
$\beta=0.4$ GeV is shown in Fig.
\ref{fig:x1835_eta1760_partial_widths}. Evidently large couplings of
$\eta(1760)$ to $\rho\rho$ and $\omega\omega$ follow from moderate
mixing, which could explain the observation of $\eta(1760)$ in the
$\rho\rho$ and $\omega\omega$ final states by the DM2 and BES
collaborations. Furthermore, we note that $\eta(1760)$ should have a
sizable branching ratio into $\pi a_2(1320)$. Therefore we urge
experimentalist to search for $\eta(1760)$ in the process
$J/\psi\rightarrow\gamma\eta(1760)\rightarrow\gamma\pi a_2(1320)$,
which is an important test to our scenario. Obviously the partial
width of $X(1835)\rightarrow\eta f_2(1270)$ is particularly small.
Taking into account the variation of the mixing angle $\phi$, we
find that $X(1835)$ may have large branching ratio into $\rho\rho$,
$\pi a_2(1320)$ and $KK^*$ final states under the assignment of
$\eta'(3{^1}{S}{_0})$ $q\bar{q}$ meson, experimental search of
$X(1835)$ in these modes is suggested.

\begin{figure}[hptb]
\begin{center}
\includegraphics*[scale=.55]{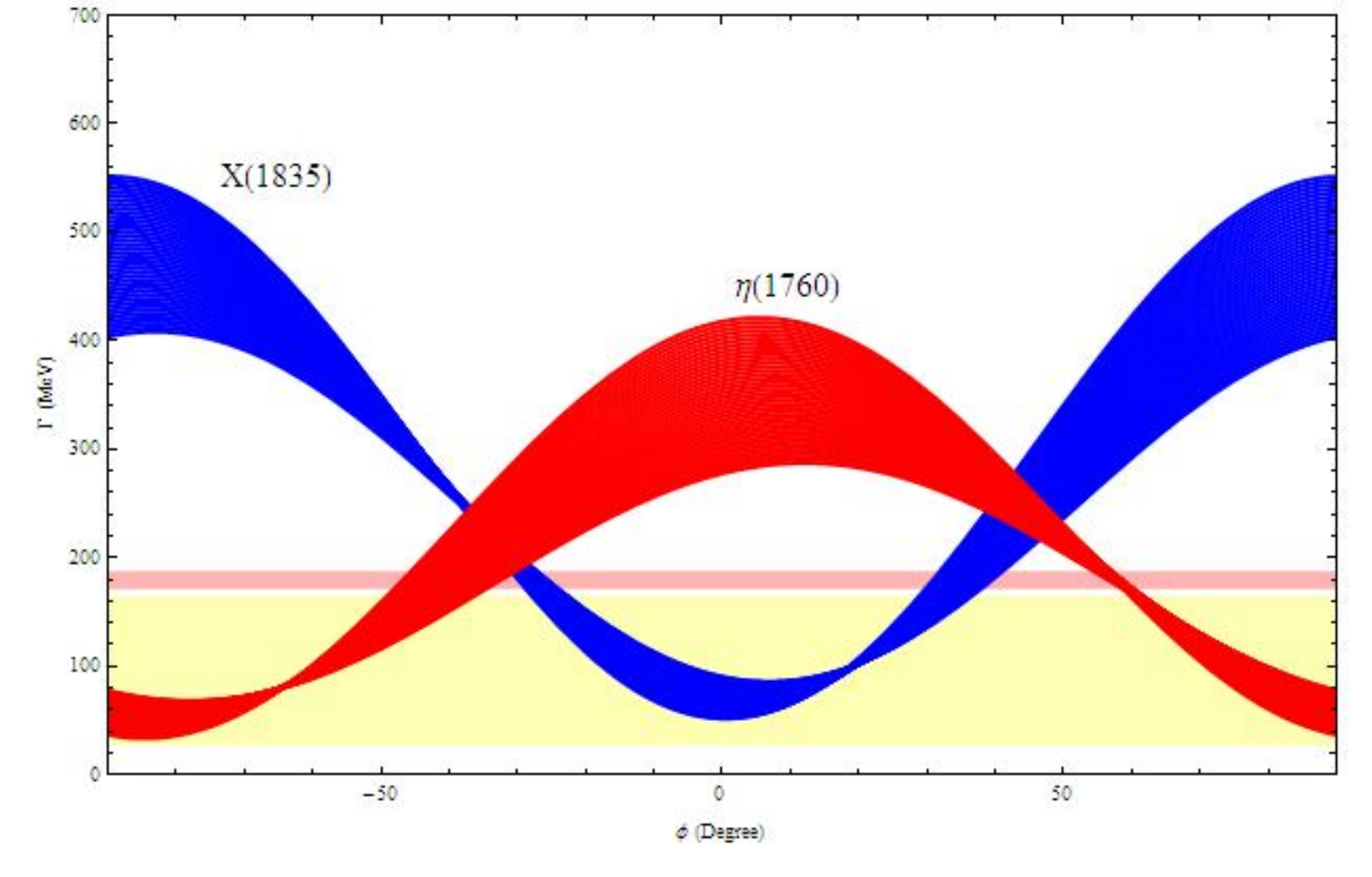}
\caption{\label{fig:eta1760-x1835}Total decay widths of $\eta(1760)$
and $X(1835)$ as functions of the mixing angle, where the harmonic
oscillator parameter $\beta$ varies from 0.35 GeV to 0.45 GeV. The
horizontal yellow and pink bands denote the experimental errors of
$\eta(1760)$ and $X(1835)$ widths, where the mass and width of
$\eta(1760)$ is taken to be the world average.}
\end{center}
\end{figure}

\begin{figure}[hptb]
\begin{center}
\begin{tabular}{cc}
\includegraphics[height=50mm]{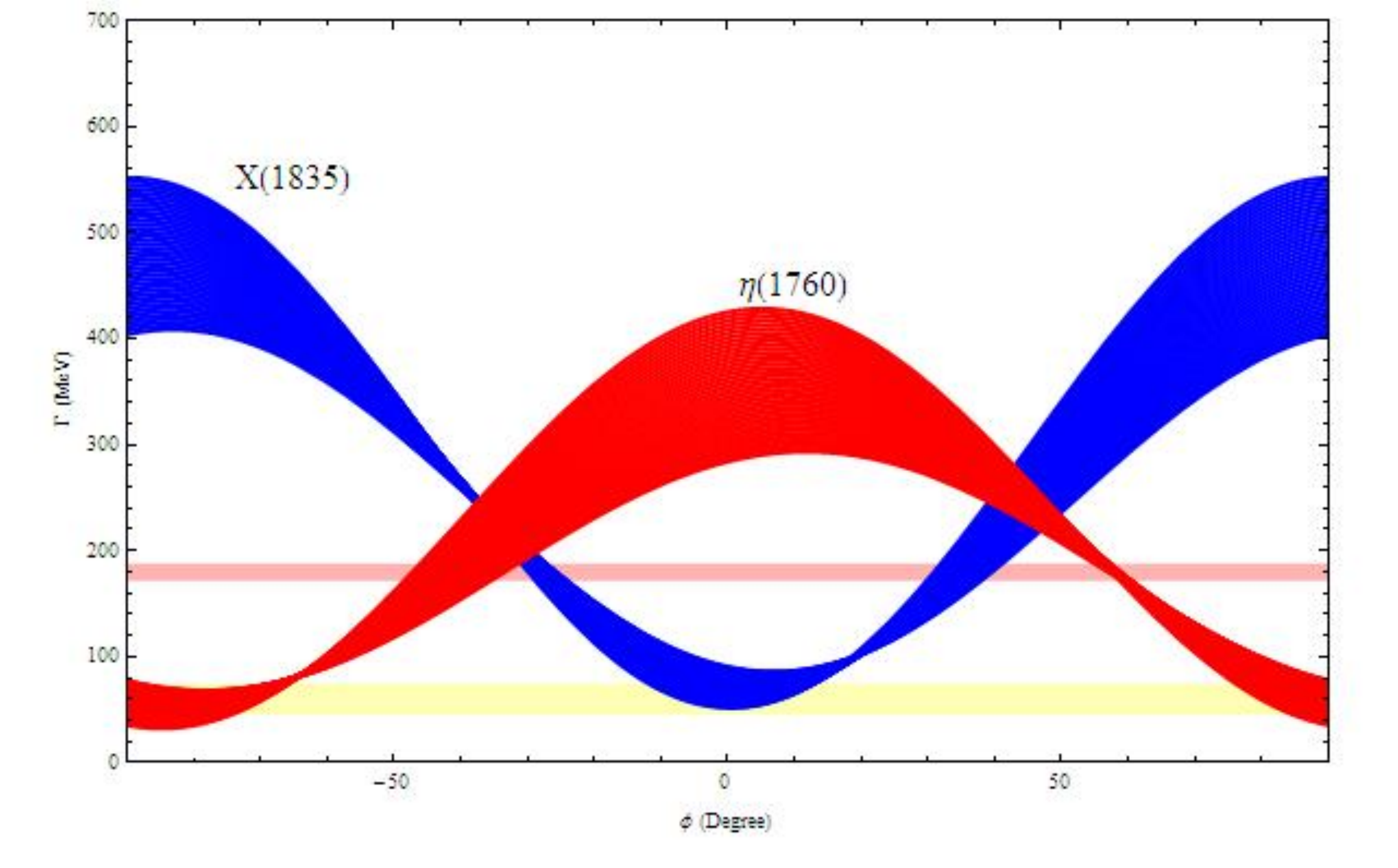}&\hspace{1cm}\includegraphics[height=50mm]{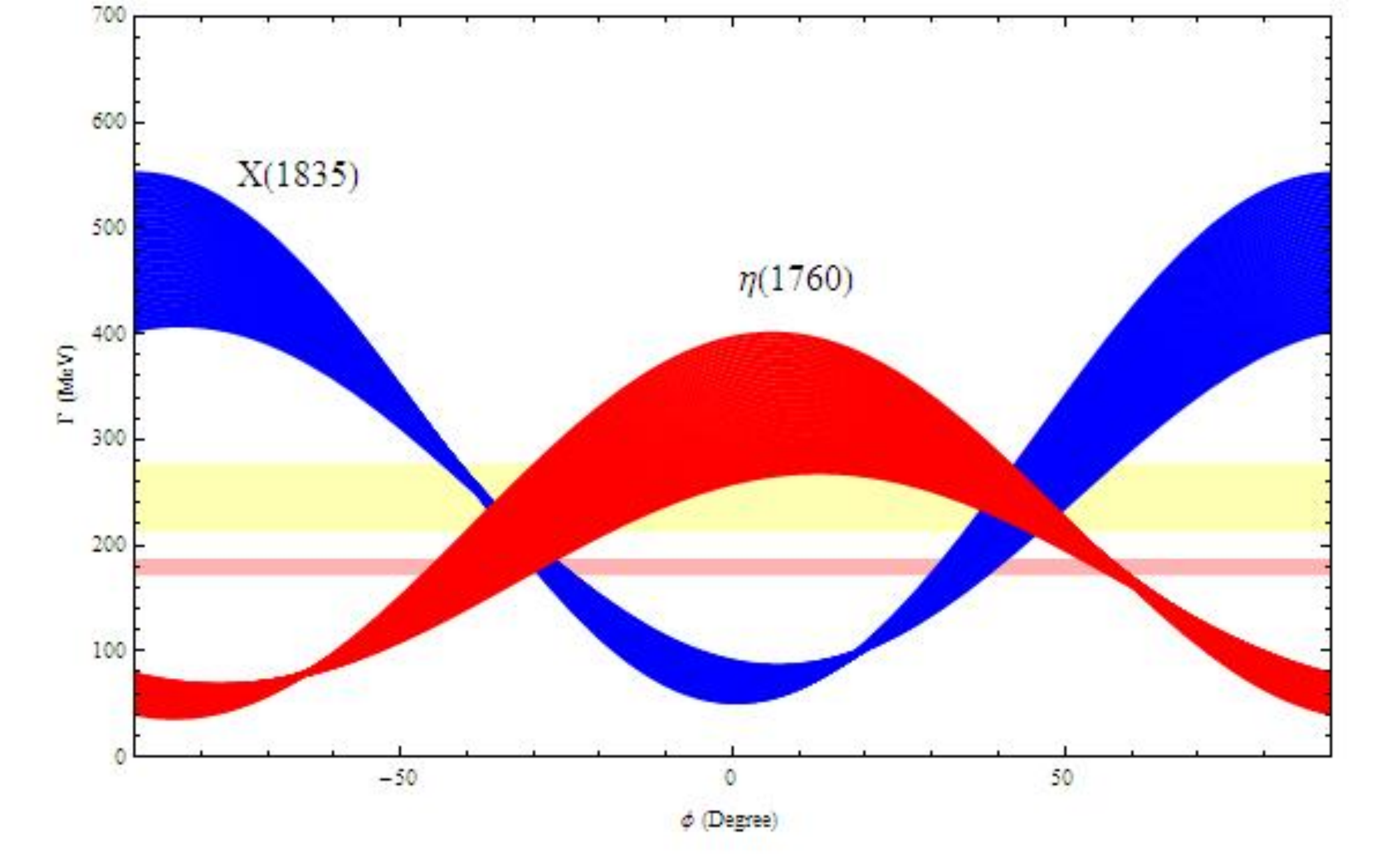}\\
(a)&(b)
\end{tabular}
\caption{\label{fig:x1835_eta1760_partial_widths_DM2_BES} The same
as Fig.\ref{fig:eta1760-x1835}, where the parameters of $\eta(1760)$
are chosen to be the DM2 and BES collaborations measurements in the
left and right figures, respectively.}
\end{center}
\end{figure}

We note that the mixing angle appearing in the $\eta(1760)$ and
$X(1835)$ flavor wavefunction is the same, so that a large number of
decays are correlated, as is demonstrated in Table \ref{tab:eta1760}
and Table \ref{tab:x1835}. It is essential to investigate whether
there exists certain region of mixing angle $\phi$ so that the
predicted widths of $\eta(1760)$ and $X(1835)$ agree with the
experimental observations within errors. Since the masses of
$\eta(1760)$ and $X(1835)$ are measured precisely enough, their
central values are used, the harmonic oscillator parameter $\beta$
is allowed to vary in the range of $0.35\sim0.45$ GeV, the total
decay widths of $\eta(1760)$ and $X(1835)$ as functions of the
mixing angle are shown in Fig. \ref{fig:eta1760-x1835}. Obviously we
see that there is not a value of $\phi$ so that the resulting widths
of both $\eta(1760)$ and $X(1835)$ lie in the experimentally allowed
range. The same conclusion is reached for the $\eta(1760)$
parameters measured by the DM2 collaboration, as is obvious from
Fig.\ref{fig:x1835_eta1760_partial_widths_DM2_BES}a. It seems
unappropriate to identify $\eta(1760)$ and $X(1835)$ as the second
radial excitation of $\eta$ and $\eta'$ simultaneously. However, if
we take the $\eta(1760)$ mass and width to be the BES measurement,
the corresponding decay widths are shown in
Fig.\ref{fig:x1835_eta1760_partial_widths_DM2_BES}b, we find that
the theoretical widths of $\eta(1760)$ and $X(1835)$ could be
consistent with experimental data for the mixing angle $\phi$ in the
range $-31^{\circ}\sim-24^{\circ}$ or $30^{\circ}\sim40^{\circ}$.
Therefore experimentally resolving the inconsistence between the DM2
and the BES collaboration results for $\eta(1760)$ is important to
understand $X(1835)$. Remembering that $X(1835)$ is close to the
threshold of proton and antiproton (i.e., $p\bar{p}$), "dressing" of
the $q\bar{q}$ singlet meson $\eta'(3{^1}{S}{_0})$ with two
$q\bar{q}$ pair can create nucleon-antinucleon, and final state
interactions enhance the probability of this transition. In this
way, the $\eta'(3{^1}{S}{_0})$ meson can mix with the $p\bar{p}$
final state and its wave function develops a sizable $p\bar{p}$
component. As a result, $X(1835)$ could be a mixture of
$\eta'(3{^1}{S}{_0})$ and $p\bar{p}$ molecule, then all experimental
facts related to $X(1835)$ could be understood qualitatively. To
shed light on the nature of $X(1835)$, a coupled channel analysis
necessary, this topic is beyond the scope of the present work.

\subsection{Decays of $X(2120)$ and $X(2370)$}
\begin{figure}[hptb]
\begin{center}
\begin{tabular}{cc}
\includegraphics[height=50mm]{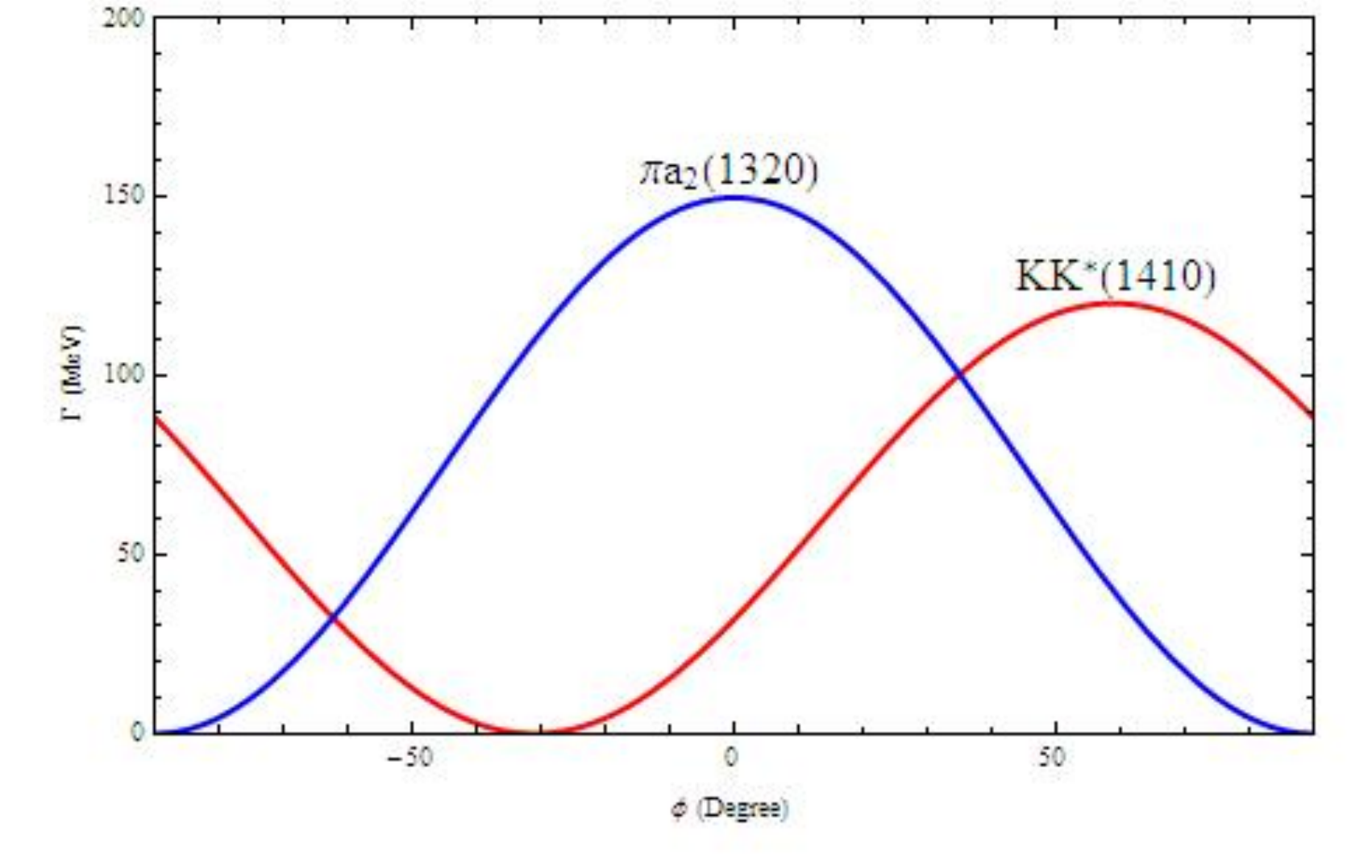}&\hspace{1cm}\includegraphics[height=50mm]{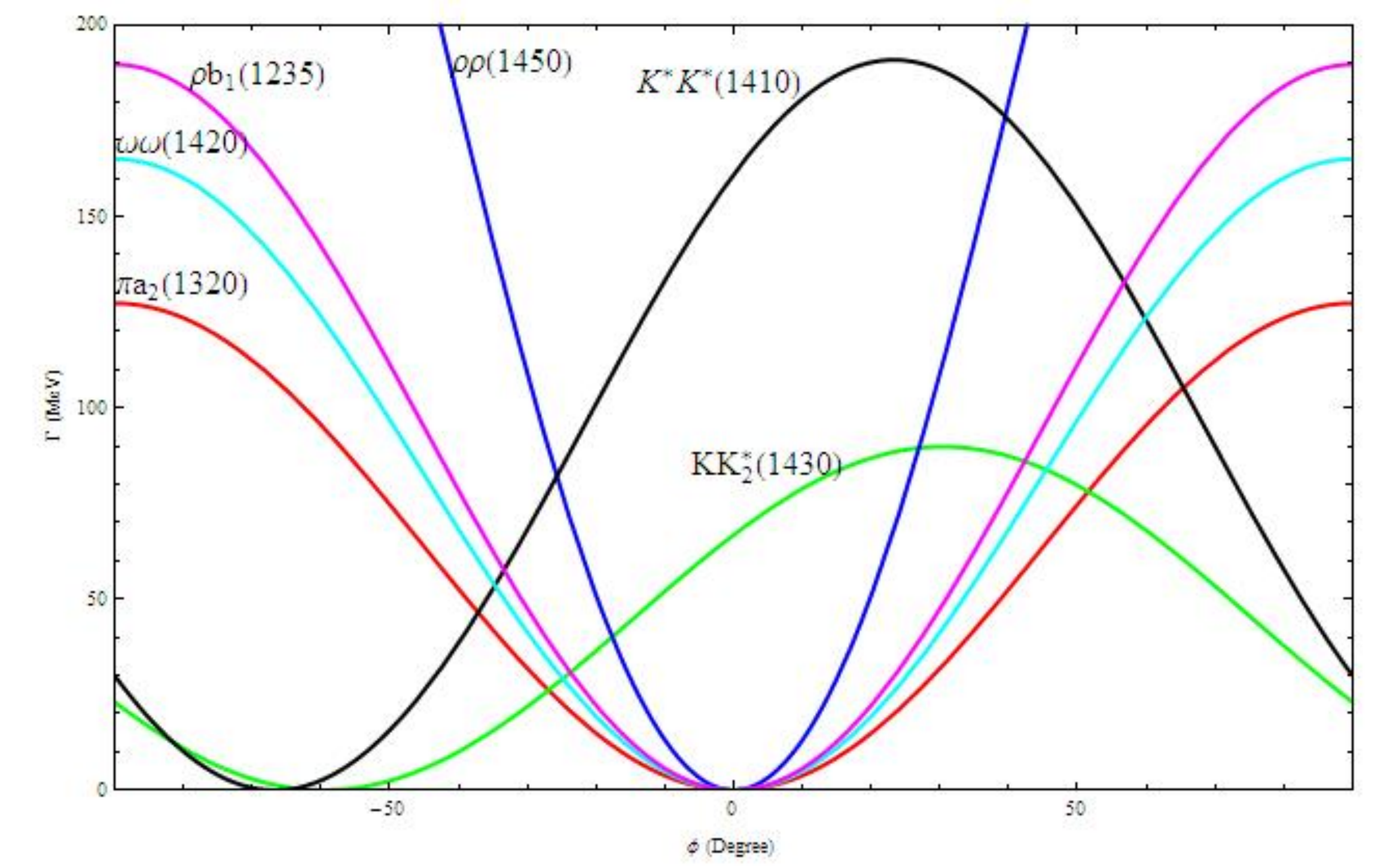}\\
(a)&(b)
\end{tabular}
\caption{\label{fig:x2120_x2370_partial_widths} Partial decay widths
of the leading decay modes of $X(2120)$ and $X(2370)$ vs. the flavor
mixing angle $\phi$, the left figure for $X(2120)$, and the right
figure for $X(2370)$.}
\end{center}
\end{figure}

Under the assignment of $\eta(4{^1}{S}{_0})$ and
$\eta'(4{^1}{S}{_0})$ $q\bar{q}$ mesons, the decay amplitudes and
partial widths of $X(2120)$ and $X(2370)$ in terms of the general
mixing angles are shown in Table \ref{tab:x2120} and Table
\ref{tab:x2370} respectively. Since $X(2120)$ and $X(2370)$ have
larger masses, many strong decay modes are allowable. $X(2120)$ has
large partial widths to $\pi a_2(1320)$ and $KK^*(1410)$, and the
main decay modes of $X(2370)$ are $\rho\rho(1450)$, $\rho
b_1(1235)$, $\omega\omega(1420)$, $\pi a_2(1320)$, $K^*K^*(1410)$
and $KK_2^*(1430)$, the corresponding partial widths as functions of
the flavor mixing angle $\phi$ are shown in
Fig.\ref{fig:x2120_x2370_partial_widths}. It is obvious that the
modes $\pi a_2(1320)$ and $KK^*(1410)$ are important to the search
for $X(2120)$, this is because that if the signal of $X(2120)$ is
accidently suppressed in one mode, it should be evident in the
other. The same is true for the $X(2370)$ decay modes
$\rho\rho(1450)$ and $K^*K^*(1410)$. We note that the branching
ratios of the $KK^*$ and $\rho\rho$ modes in both $X(2120)$ and
$X(2370)$ decays are predicted to be smaller, despite their larger
phase space, as they are accidentally near the node in the $^3P_0$
decay amplitude for the physical masses and $\beta=0.4$ GeV. The
$X(2120)$ decay modes $\rho b_1(1235)$ and $\omega h_1(1170)$ are
interesting because the two subamplitudes $^1S_0$ and $^5D_0$ are
comparable and individually proportional to $\cos\phi$, thus the
$D/S$ amplitude ratio is independent of the mixing angle $\phi$. The
measurement of $\rho b_1(1235)$ and $\omega h_1(1170)$ subamplitudes
directly access $\cos\phi$, although these modes may be too weak to
allow this measurement. Similarly $X(2370)$ can decay into $\rho
b_1(1235)$, $\omega h_1(1170)$, $K^*K_1(1273)$, $K^*K_1(1402)$ in
both S-wave and D-wave, and the $D/S$ ratio for the latter two modes
strongly depends on the flavor mixing angle.

\begin{figure}[hptb]
\begin{center}
\includegraphics*[scale=.65]{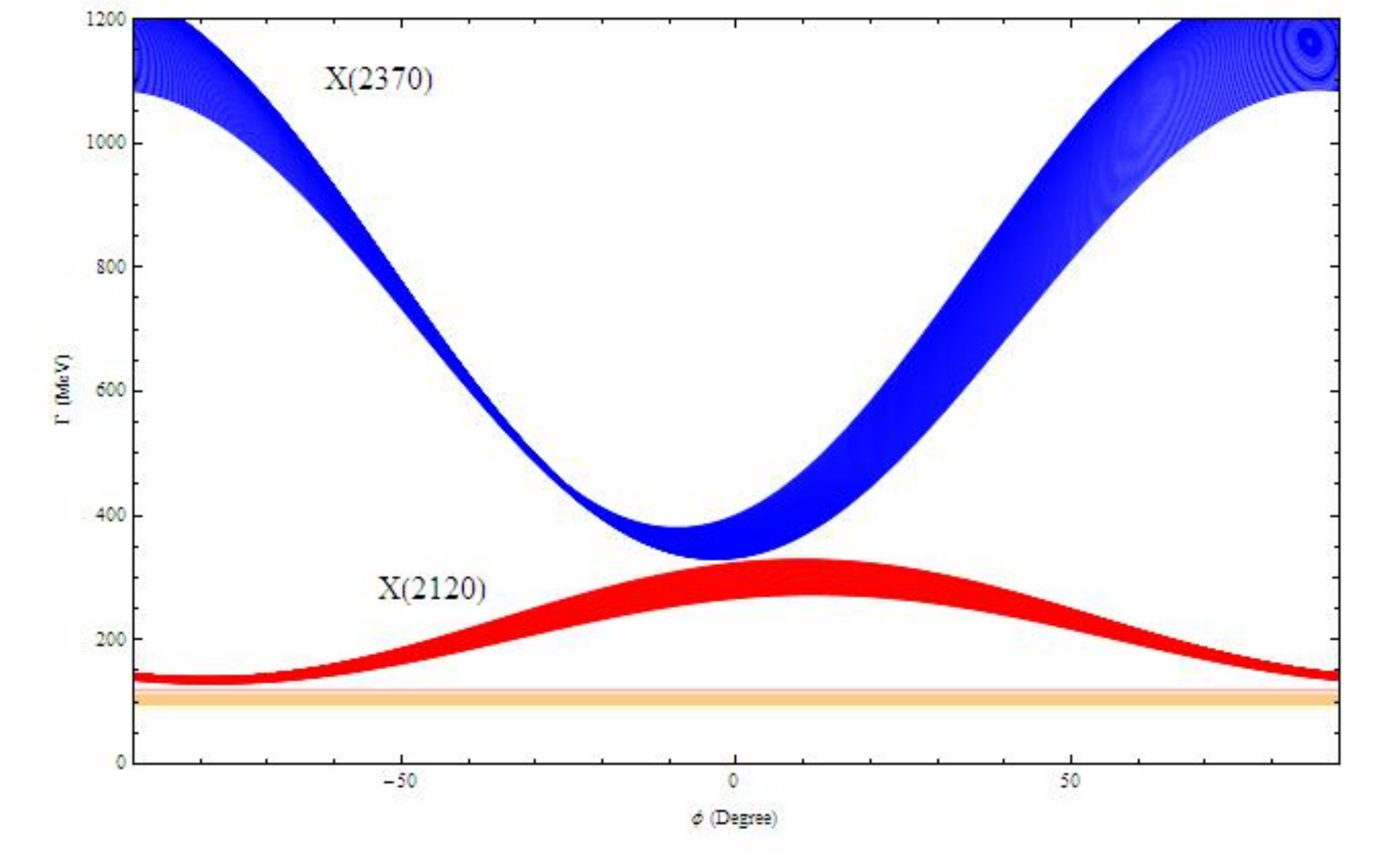}
\caption{\label{fig:x2120-x2370}Total decay widths of $X(2120)$ and
$X(2370)$ vs. the flavor mixing angle $\phi$, where the harmonic
oscillator parameter $\beta$ varies from 0.35 GeV to 0.45 GeV. The
horizontal yellow and pink bands represent the widths of $X(2120)$
and $X(2370)$ fitted by BES collaboration respectively, which are
close to each other.}
\end{center}
\end{figure}

For the harmonic oscillator parameter $\beta$ in the range of 0.35
GeV$\sim$0.45 GeV, the total widths of $X(2120)$ and $X(2370)$
against the flavor mixing angle $\phi$ is displayed in Fig.
\ref{fig:x2120-x2370}. Since $X(2370)$ has many decay modes, its
width is predicted to be larger than 300 MeV. Even if the width is
overestimated by a factor of 2, it is still larger than the measured
value. Obviously there doesn't exist appropriate value of the mixing
angle so that the theoretically predicted widths of $X(2120)$ and
$X(2370)$ lie in the experimentally allowed range. Therefore it
seems very unlikely that $X(2120)$ and $X(2370)$ can be understood
as the third radial excitation of $\eta$ and $\eta'$ simultaneously.
The lattice QCD simulations predict the $0^{-+}$ glueball is about
2.3$\sim$2.6 GeV \cite{Morningstar:1999rf}, it would mix with the
nearby pseudoscalar isoscalar mesons. Consequently $X(2370)$ may be
a mixture of $\eta'(4{^1}{S}{_0})$ and glueball, if its quantum
numbers turn out to be $J^{PC}=0^{-+}$ in future. To understand the
nature of $X(2370)$, partial wave analysis is important.

\section{Summary and discussions }
In this work, we investigate whether the resonances $X(1835)$,
$X(2120)$ and $X(2370)$ newly observed by the BES collaboration
could be conventional $q\bar{q}$ mesons. If they are indeed
canonical pseudoscalar mesons, the natural assignments are
$\eta(1760)$ and $X(1835)$ as the second radial excitation of $\eta$
and $\eta'$ respectively, and $X(2120)$ and $X(2370)$ as the third
radial excitation of $\eta$ and $\eta'$. To do so we calculate all
kinematically allowed two-body strong decays of
$\eta(3{^1}{S}{_0})$, $\eta'(3{^1}{S}{_0})$, $\eta(4{^1}{S}{_0})$
and $\eta'(4{^1}{S}{_0})$ states within the framework of
${^3}{P}{_0}$ model.

The decay amplitudes and widths turn out to be strongly dependent on
the flavor mixing angle. If the mass and width of $\eta(1760)$ are
chosen to be the world average listed in PDG or the DM2 measurement,
we can not find proper value of the mixing angle so that both the
theoretically predicted widths of $\eta(1760)$ and $X(1835)$ lie in
the experimentally allowed range. However, if the BES results for
$\eta(1760)$ are taken to be true, the theoretical predictions could
be consistent with the experimental data within error for the flavor
mixing angle $\phi$ in the range of $-31^{\circ}\sim-24^{\circ}$ or
$30^{\circ}\sim40^{\circ}$. Further experimental study of
$\eta(1760)$ is important to understand the nature of $X(1835)$.
Since the $\eta'(3{^1}{S}{_0})$ $q\bar{q}$ meson would mix with
$p\bar{p}$ due to the "dressing" effect and final state interaction,
we suggest $X(1835)$ is the mixture of $\eta'(3{^1}{S}{_0})$ and
$p\bar{p}$ molecule, then we can naturally understand all the
observations associated with $X(1835)$.

Under the assignment of $X(2120)$ and $X(2370)$ as
$\eta(4{^1}{S}{_0})$ and $\eta'(4{^1}{S}{_0})$ $q\bar{q}$ mesons,
$X(2120)$ dominantly decays into $\pi a_2(1320)$ and $KK^*(1410)$,
the modes $KK^*$ and $\rho\rho$ modes are suppressed by the decay
amplitude node. $X(2370)$ is predicted to be rather broad (i.e., its
width should be larger than 300 MeV), so it is unlikely that
$X(2120)$ and $X(2370)$ can be understood as the third radial
excitation of $\eta$ and $\eta'$ simultaneously. Since $X(2370)$ is
close to the $0^{-+}$ glueball 2.3$\sim$2.6 GeV predicted by lattice
QCD, we suggest it may be a mixture of $\eta'(4{^1}{S}{_0})$ meson
and glueball, if its quantum numbers are determined to be
$J^{PC}=0^{-+}$ by future experiments.

\begin{acknowledgments}
We are grateful to Prof. Dao-Neng Gao for stimulating discussions.
This work is supported by the National Natural Science Foundation of
China under Grant No.10905053, No. 10975128, Chinese Academy
KJCX2-YW-N29 and the 973 project with Grant No. 2009CB825200.
Jia-Feng Liu is supported in part by the National Natural Science
Foundation of China under Grant No.10775124.
\end{acknowledgments}

\newpage

\begin{table}[hptb]
\begin{center}
\begin{tabular}{||c|c|c|c|c||}
\hline\hline Generic Decay&~~~~~Example                      &
$~~~~~I_{flavor}(a)$  &~~~ $I_{flavor}(b)$ & ~~~${\cal F}$
\\\hline\hline
$X_0\rightarrow(n\bar{s})(s\bar{n})$ & \small
$\ba{c} X_{0}\rightarrow{K}^{+}+K^{-}\\
X_{0}\rightarrow{K^{*}}^{+}+K^{-} \ea $ & \small
$\ba{c}0\\
0 \ea $ & \small $\ba{c}-1/\sqrt{2}\\ -1/\sqrt{2} \ea $ & \small
$\ba{c}2 \\4\ea $
\\
\hline

$X_s\rightarrow(n\bar{s})(s\bar{n})$ & \small
$\ba{c} X_{s}\rightarrow{K}^{+}+K^{-}\\
X_{s}\rightarrow{K^{*}}^{+}+K^{-} \ea $ & \small
$\ba{c}-1\\
-1 \ea $ & \small $\ba{c}0\\ 0 \ea $ & \small $\ba{c}2
\\4\ea $
\\
\hline

$X_0\rightarrow(u\bar{d})(d\bar{u})$

& \small $\ba{c}X_{0}\rightarrow \pi^{+}+a^{-}\\X_{0}\rightarrow
\pi^{+}+\pi^{-}\\X_{0}\rightarrow
\rho^{+}+\rho^{-}\\X_{0}\rightarrow \rho^{+}+\rho^{-}(1450)\ea $

& \small
$\ba{c}-1/\sqrt{2}\\-1/\sqrt{2}\\-1/\sqrt{2}\\-1/\sqrt{2}\\\ea $

& \small
$\ba{c}-1/\sqrt{2}\\-1/\sqrt{2}\\-1/\sqrt{2}\\-1/\sqrt{2}\\\ea $

& \small $\ba{c}3 \\3/2 \\3/2 \\3 \ea $
\\
\hline

$X_s\rightarrow(u\bar{d})(d\bar{u})$

& \small $\ba{c}X_{s}\rightarrow \pi^{+}+a^{-}\\X_{s}\rightarrow
\pi^{+}+\pi^{-}\\X_{s}\rightarrow
\rho^{+}+\rho^{-}\\X_{s}\rightarrow \rho^{+}+\rho^{-}(1450)\ea $

& \small $\ba{c}0\\0\\0\\0\\\ea $

& \small $\ba{c}0\\0\\0\\0\\\ea $

& \small $\ba{c}0\\0\\0\\0\\\ea $

\\
\hline

$X_0\rightarrow(\frac{u\bar{u}+d\bar{d}}{\sqrt{2}})(\frac{u\bar{u}+d\bar{d}}{\sqrt{2}})$

& \small $\ba{c}X_{0}\rightarrow \eta_{0}+f_{0}\\X_{0}\rightarrow
\omega+\omega\\\ea $

& \small $\ba{c}1/\sqrt{2}\\1/\sqrt{2}\\\ea $

& \small $\ba{c}1/\sqrt{2}\\1/\sqrt{2}\\\ea $

& \small $\ba{c}1\\1/2\\\ea $
\\
\hline
$X_s\rightarrow(\frac{u\bar{u}+d\bar{d}}{\sqrt{2}})(\frac{u\bar{u}+d\bar{d}}{\sqrt{2}})$& $X_{s}\rightarrow \eta_{0}+f_{0}$   & $0$              & $0$             &   $0$ \\
\hline
$X_0\rightarrow(s\bar{s})(\frac{u\bar{u}+d\bar{d}}{\sqrt{2}})$& $X_{0}\rightarrow \eta_{s}+f_{0}$   & $0$              & $0$             &   $0$ \\
\hline
$X_s\rightarrow(s\bar{s})(\frac{u\bar{u}+d\bar{d}}{\sqrt{2}})$& $X_{s}\rightarrow \eta_{s}+f_{0}$   & $0$              & $0$             &   $0$ \\
\hline
$X_0\rightarrow(\frac{u\bar{u}+d\bar{d}}{\sqrt{2}})(s\bar{s})$& $X_{0}\rightarrow \eta_{0}+f_{s}$   & $0$              & $0$             &   $0$ \\
\hline
$X_s\rightarrow(\frac{u\bar{u}+d\bar{d}}{\sqrt{2}})(s\bar{s})$& $X_{s}\rightarrow \eta_{0}+f_{s}$   & $0$              & $0$             &   $0$ \\
\hline
$X_0\rightarrow(s\bar{s})(s\bar{s})$& $X_{0}\rightarrow \eta_{s}+f_{s}$   & $0$              & $0$             &   $0$ \\
\hline $X_s\rightarrow(s\bar{s})(s\bar{s})$

& \small $\ba{c}X_{s}\rightarrow \eta_{s}+f_{s}\\X_{s}\rightarrow
\phi+\phi \\\ea $

& \small $\ba{c}1\\1\\\ea $

& \small $\ba{c}1\\1\\\ea $

& \small $\ba{c}1\\1/2\\\ea $
\\

\hline\hline
\end{tabular}
\caption{\label{tab:flavor}Relevant flavor weight factors for $\eta$
and $\eta'$ excitation decays, where $|X_{0}\rangle=|u\bar u+d\bar
d\rangle/\sqrt{2}$ and $|X_{s}\rangle=|s\bar s\rangle$,
$(n\bar{s})=(u\bar{s})$ or $(d\bar{s})$ for $n$ being up and down
quark respectively. $(n\bar{n}')_{I=1}=$$(u\bar{d})$,
$[(u\bar{u})-(d\bar{d})]/\sqrt{2}$ and $(d\bar{u})$,
$(n\bar{n})_{I=0}=[(u\bar{u})+(d\bar{d})]/\sqrt{2}$.}
\end{center}
\end{table}

\begin{table}[hptb]
\begin{center}
\renewcommand{\arraystretch}{0.5}
\begin{tabular}{||c|c|c||} \hline\hline
\multicolumn{3}{||c||}{$\eta(1760)=\cos\phi_1|u\bar{u}+d\bar{d}\rangle/\sqrt{2}-\sin\phi_1|s\bar{s}\rangle$}\\\hline
Modes & $\Gamma({\rm  MeV})$ & \multicolumn{1}{c||}{Amps.({$\rm GeV^{-1/2}$})} \\
\hline\hline
$K K^*$ & $30.85c^2+89.25cs+64.56s^2$  & ${\cal M}_{11}=0.074c+0.11s$ \\
\hline
$\rho \rho$ & $155.36c^2$  &  ${\cal M}_{11}=0.30c$ \\
\hline
$\omega \omega$ & $49.50c^2$   & ${\cal M}_{11}=-0.30 c$ \\
\hline
$\pi a_{0}(1450)$ & $44.36c^2$  & ${\cal M}_{00}=-0.22 c$ \\
\hline
$\pi a_{2}(1320)$ & $60.93c^2$   & ${\cal M}_{22}=-0.17 c$ \\
\hline
Total & $341.00c^2+89.25cs+64.56s^2$ & $\qquad$  \\
\hline\hline
\end{tabular}
\caption{\label{tab:eta1760}Partial widths of $\eta(1760)$ as the
second radial excitation of $\eta$, where $\phi_1$ is the flavor
mixing angle, $s\equiv\sin\phi_1$ and $c\equiv\cos\phi_1$. Note that
a factor of $i$ has been suppressed in all odd partial wave
amplitudes.}
\renewcommand{\arraystretch}{0.5}
\end{center}
\end{table}

\begin{table}[hptb]
\begin{center}
\renewcommand{\arraystretch}{0.5}
\begin{tabular}{||c|c|c||} \hline\hline
\multicolumn{3}{||c||}{$X(1835)=\sin\phi_1|u\bar{u}+d\bar{d}\rangle/\sqrt{2}+\cos\phi_1|s\bar{s}\rangle$}\\\hline
Modes & $\Gamma({\rm  MeV})$ & \multicolumn{1}{c||}{Amps.({$\rm GeV^{-1/2}$})}  \\
\hline\hline
$K K^*$ & $43.18c^2-79.34cs+36.45s^2$  & ${\cal M}_{11}=-0.080c+0.074s$ \\
\hline
$\rho \rho$ & $188.24s^2$  &  ${\cal M}_{11}=0.30s$ \\
\hline
$K^* K^*$ & $29.84c^2+23.08cs+4.46s^2$   & ${\cal M}_{11}=0.16c+0.060s$ \\
\hline
$\omega \omega$ & $62.23s^2$   & ${\cal M}_{11}=-0.30 s$ \\
\hline
$\pi a_{0}(1450)$ & $47.85s^2$  & ${\cal M}_{00}=-0.18 s$ \\
\hline
$\pi a_{2}(1320)$ & $136.48s^2$   & ${\cal M}_{22}=-0.22 s$ \\
\hline
$\eta f_{2}(1270)$ & $0.051s^2$   & ${\cal M}_{22}=0.014s$ \\
\hline
Total & $73.02c^2-56.26cs+475.76s^2$ & $\qquad$  \\
\hline\hline
\end{tabular}
\caption{\label{tab:x1835}Partial widths of $X(1835)$ as the second
radial excitation of $\eta'$, where $\phi_1$ is the mixing angle,
$s\equiv\sin\phi_1$ and $c\equiv\cos\phi_1$, and the factor of $i$
has been suppressed in all odd partial wave amplitudes.}
\renewcommand{\arraystretch}{0.5}
\end{center}
\end{table}

\begin{table}[hptb]
\begin{center}
\renewcommand{\arraystretch}{0.5}
\begin{tabular}{||c|c|c||}
\hline\hline
\multicolumn{3}{||c||}{$X(2120)=\cos\phi_2|u\bar{u}+d\bar{d}\rangle/\sqrt{2}-\sin\phi_2|s\bar{s}\rangle$}\\\hline
Mode & $\Gamma({\rm  MeV})$ & \multicolumn{1}{c||}{Amps.({$\rm GeV^{-1/2}$})}  \\
\hline\hline
$K K^*$ & $2.39c^2-9.26cs+8.98s^2$ & ${\cal M}_{11}=-0.015c+0.029s$  \\
\hline
$KK^*(1410)$ & $31.90c^2+106.14cs+88.29s^2$ & ${\cal M}_{11}=0.082c+0.14s$  \\
\hline
$\pi a_{0}(1450)$ & $0.013c^2$ & ${\cal M}_{00}=-0.0018 c$  \\
\hline
$KK_{0}^*(1430)$ & $2.98c^2-3.70cs+1.15s^2$ & ${\cal M}_{00}=0.025c-0.016s$ \\
\hline
$\eta f_{0}(1370)$ & $1.61c^{2}$ & ${\cal M}_{00}=-0.036c$  \\
\hline
$\pi a_{2}(1320)$ & $149.68c^{2}$ & ${\cal M}_{22}=0.16 c$  \\
\hline
$K K_{2}^*(1430)$ & $3.72c^2-23.98cs+38.63s^2$ & ${\cal M}_{22}=0.028 c-0.092 s$  \\
\hline
$\eta f_{2}(1270)$ & $21.55c^2$ & ${\cal M}_{22}=-0.12 c$  \\
\hline
$\eta f_{2}^{\prime}(1525)$ & $0.25s^2$ & ${\cal M}_{22}=-0.021 s$  \\
\hline
$\rho \rho$ & $1.09c^{2}$ & ${\cal M}_{11}=-0.017c$  \\
\hline
$K^* K^*$ & $5.60c^2-6.20cs+1.71s^2$ & ${\cal M}_{11}=-0.038c+0.021s$  \\
\hline
$\phi \phi$ & $4.67s^{2}$ & ${\cal M}_{11}=-0.097s$  \\
\hline
$\omega \omega$ & $0.52c^{2}$ & ${\cal M}_{11}=0.021c$  \\
\hline

$\rho b_{1}(1235)$ & $50.80c^{2}$
& \small $\ba{c} {\cal M}_{00}=0.082 c\\
{\cal M}_{22}=0.093 c \ea $
\\
\hline

$\omega h_{1}(1170))$ & $22.75c^{2}$ & \small
$\ba{c}{\cal M}_{00}=-0.051 c\\
{\cal M}_{22}=-0.12 c \ea $
\\
\hline
Total & $294.60c^2+63.00cs+143.69s^2$ & $\qquad$  \\
\hline\hline
\end{tabular}
\renewcommand{\arraystretch}{0.5}
\end{center}
\caption{\label{tab:x2120} Partial widths of $X(2120)$ as the third
radial excitation of $\eta$, where $s\equiv\sin\phi_2$ and
$c\equiv\cos\phi_2$, and the factor of $i$ has been suppressed in
all odd partial wave amplitudes.}
\end{table}

\begin{table}[hptb]
\begin{center}
\renewcommand{\arraystretch}{0.5}
\begin{tabular}{||c|c|c||}
\hline\hline
\multicolumn{3}{||c||}{$X(2370)=\sin\phi_2|u\bar{u}+d\bar{d}\rangle/\sqrt{2}+\cos\phi_2|s\bar{s}\rangle$}\\\hline
Modes & $\Gamma({\rm  MeV})$ & \multicolumn{1}{c||}{Amps.({$\rm GeV^{-1/2}$})} \\
\hline\hline
$KK^*$ & $14.33c^2+1.94cs+0.066s^2$ & ${\cal M}_{11}=-0.032c-0.0022s$  \\
\hline
$K(1460)K^*$ & $17.65c^2-13.33cs+2.52s^2$ & ${\cal M}_{11}=-0.10c+0.036s$ \\
\hline
$KK^*(1410)$ & $2.64c^2-22.59cs+48.37s^2$ & ${\cal M}_{11}=-0.017c+0.075s$  \\
\hline
$\pi a_{0}(1450)$ & $14.83s^{2}$ & ${\cal M}_{00}=-0.047s$  \\
\hline
$K K_{0}^*(1430)$ & $9.41c^2+0.15cs+0.00064s^2$ & ${\cal M}_{00}=-0.033c-0.00027s$ \\
\hline
$\eta f_{0}(1370)$ & $1.74s^{2}$ & ${\cal M}_{00}=0.028s$  \\
\hline
$\eta^{\prime} f_{0}(1370)$ & $5.61s^{2}$ & ${\cal M}_{00}=-0.084 s$  \\
\hline
$\eta f_{0}(1710)$ & $2.61c^{2}$ & ${\cal M}_{00}=0.055c$  \\
\hline
$\pi a_{2}(1320)$ & $127.35s^{2}$ & ${\cal M}_{22}=0.12s$  \\
\hline
$K K_{2}^*(1430)$ & $66.79c^2+78.46cs+23.04s^2$ & ${\cal M}_{22}=0.089c+0.052s$ \\
\hline
$\eta f_{2}(1270)$ & $36.25s^2$ & ${\cal M}_{22}=-0.12s$  \\
\hline
$\eta^{\prime} f_{2}(1270)$ & $7.50s^2$ & ${\cal M}_{22}=-0.072s$ \\
\hline
$\eta f_{2}^{\prime}(1525)$ & $11.60c^2$ & ${\cal M}_{22}=0.083c$ \\
\hline
$K K^*(1680)$ & $9.30c^2-6.24cs+1.05s^2$ & ${\cal M}_{11}=-0.047c+0.016s$  \\
\hline
$K K_{3}^*(1780)$ & $2.12c^2-0.72cs+0.061s^2$ & ${\cal M}_{33}=-0.026c+0.0044s$  \\
\hline
$\rho \rho$ & $12.56s^{2}$ & ${\cal M}_{11}=0.050s$  \\
\hline
$K^* K^*$ & $9.15c^2-10.41cs+2.96s^2$ & ${\cal M}_{11}=0.040c-0.023s$  \\
\hline
$\phi \phi$ & $3.88c^{2}$ & ${\cal M}_{11}=0.059c$  \\
\hline
$\omega \omega$ & $3.84s^{2}$ & ${\cal M}_{11}=-0.048s$  \\
\hline
$\rho \rho(1450)$ & $435.60s^{2}$ & ${\cal M}_{11}=0.34s$  \\
\hline
$K^* K^*(1410)$ & $161.05c^2+138.87cs+29.94s^2$ & ${\cal M}_{11}=0.21c+0.089s$  \\
\hline
$\omega \omega(1420)$ & $165.10s^{2}$ & ${\cal M}_{11}=-0.33s$  \\
\hline $\rho b_{1}(1235)$ & $189.78s^{2}$ & \small $\ba{c} {\cal
M}_{00}=-0.028s
\\
{\cal M}_{22}=0.17s \ea $
\\
\hline $K^* K_{1}(1273)$ & $12.76c^2+26.21cs+14.04s^2$ & \small
$\ba{c} {\cal M}_{00}=-0.0087c
\\
{\cal M}_{22}=0.042c+0.045s \ea $
\\
\hline $\omega h_{1}(1170)$ & $68.36s^{2}$ & \small $\ba{c}{\cal
M}_{00}=0.038s
\\
{\cal M}_{22}=-0.16s \ea $
\\
\hline
$K^* K_{1}(1402)$ & $24.78c^2-18.75cs+17.85s^2$ & \small $\ba{c}
{\cal M}_{00}=0.034c-0.066s
\\
{\cal M}_{22}=-0.070c \ea $
\\
\hline
$K^* K_{2}^*(1430)$ & $9.01c^2-4.51cs+0.56s^2$ & ${\cal M}_{22}=-0.052c+0.013s$ \\
\hline
Total & $357.08c^2+169.10cs+1208.97s^2$ & \qquad  \\
\hline\hline
\end{tabular}
\renewcommand{\arraystretch}{0.5}
\end{center}
\caption{\label{tab:x2370} Partial widths of $X(2370)$ as the third
radial excitation of $\eta'$, where $s\equiv\sin\phi_2$ and
$c\equiv\cos\phi_2$, $\phi_2$ is the mixing angle between $X(2120)$
and $X(2370)$, and the factor of $i$ has been suppressed in all odd
partial wave amplitudes.}
\end{table}

\end{document}